\documentclass[grl]{agutex}

\usepackage[dvips]{graphicx}

\authorrunninghead{HATANO}

\titlerunninghead{CRITICAL SLIP DISTANCE: GRANULAR LAYERS}

\authoraddr{Takahiro Hatano, Earthquake Research Institute, 
University of Tokyo, 1-1-1 Yayoi, Bunkyo,  Tokyo 113-0032, Japan
(hatano@eri.u-tokyo.ac.jp)}

\begin{document}

%
%

\title{Scaling of the critical slip distance in granular layers}

%
%


\authors{Takahiro Hatano \altaffilmark{1}}
\altaffiltext{1}{Earthquake Research Institute, University of Tokyo, Tokyo, Japan}

%
%

\begin{abstract}
We investigate the nature of friction in granular layers by means of numerical simulation 
focusing on the critical slip distance, over which the system relaxes to a new stationary state.
Analyzing a transient process in which the sliding velocity is instantaneously changed, 
we find that the critical slip distance is proportional to the sliding velocity.
We thus define the relaxation time, which is independent of the sliding velocity.
It is found that the relaxation time is proportional to the layer thickness and 
inversely proportional to the square root of the pressure.
An evolution law for the relaxation process is proposed, which does not contain 
any length constants describing the surface geometry but the relaxation time of the bulk granular matter.
As a result, the critical slip distance is scaled with a typical length scale of a system.
It is proportional to the layer thickness in an instantaneous velocity change experiment, 
whereas it is scaled with the total slip distance in a spring-block system on granular layers.
\end{abstract}

%
%

%

\begin{article}

\section{Introduction}
A natural fault has the cataclasite core zone, along which shear deformation concentrates 
(e.g. \cite{engelder,scholz}).
Rheology of these granular matters thus provides us an important insight 
in considering the nature of friction on faults from a microscopic point of view.
Unfortunately, to this date, our understanding of the rheological properties of granular matter 
is still poor except for dilute flow to which the kinetic theory of gases can apply 
(e.g. \cite{garzo} and references therein.)
Thus, a computational approach has played a considerable role in investigating 
dense granular rheology to propose some constitutive laws for stationary shear flow 
(e.g. \cite{aharonov,mair,GDR,dacruz,pouliquen,hatano}).
However, a nonstationary state is still a frontier in the sense that we do not have 
any constitutive laws for transient processes.

The description of transient states is particularly important in the context of seismology 
because an earthquake is essentially a nonstationary process.
An important quantity is the critical slip distance, over which a fault looses its frictional strength 
with the coseismic slip \citep{ida}, because it determines the maximum acceleration 
of the seismic ground motion \citep{aki} as well as the rupture nucleation process 
(e.g. \cite{ohnaka1}).
However, regardless of its importance, we still can not explain the critical slip distance 
ranging from $10^{-1}$ to $1$ m, which is obtained by the seismic inversion \citep{ide}.
It is rather paradoxical that the critical slip distance obtained in a typical experiment 
is of the order of $10^{-5}$ m \citep{dieterich,scholtz}.
Understanding the physics that determines the critical slip distance to explain 
the wide gap between a natural fault and a laboratory is thus a central problem in seismology 
\citep{marone1,ohnaka2,ohnaka3}.

In this letter, by means of numerical simulation on sheared granular layers, 
we obtain a constitutive law that describes a nonstationary process as well as a stationary state.
Using this constitutive law together with dimensional analysis, 
we propose a new interpretation on the scale dependence of the critical slip distance.

\section{Model}
In the following we describe the computational model of granular layers.
Each grain is assumed to be sphere.
The interaction between grains is described by the discrete element method (DEM) 
\citep{cundall}, which is the standard model used in powder engineering and soil mechanics.
Consider a grain $i$ of radius $R_i$ located at ${\bf r_i}$ with the translational velocity ${\bf v_i}$ 
and the angular velocity $\bf\Omega_i$.
This grain interacts with another grain $j$ when they are in contact; i.e. $|{\bf r}_{ij}|<R_i+R_j$, 
where ${\bf r}_{ij}={\bf r}_i-{\bf r}_j$.
The interaction consists of two kinds of forces, which are normal and transverse to ${\bf r}_{ij}$, respectively.
Introducing the unit normal vector ${\bf n}_{ij}={\bf r}_{ij}/|{\bf r}_{ij}|$, 
the normal force acting on $i$, which is denoted by ${\bf F}^{(n)}_{ij}$, is given by 
$\left[k h_{ij}+\zeta{\bf n}_{ij}\cdot\dot{{\bf r}}_{ij}\right]{\bf n}_{ij}$,
where $h_{ij}=R_i+R_j-|{\bf r}_{ij}|$.
In order to define the transverse force, we utilize the relative tangential velocity ${\bf v}^{(t)}_{ij}$ 
defined by $\dot{{\bf r}}_{ij} - {\bf n}_{ij}\cdot\dot{{\bf r}}_{ij} + (R_i{\bf\Omega}_i+R_j{\bf\Omega}_j)/(R_i+R_j)\times{\bf r}_{ij}$ 
and introduce the relative tangential displacement vector 
${\bf \Delta}^{(t)}_{ij}=\int_{\rm roll} dt{\bf v}^{(t)}_{ij}$.
The subscript in the integral indicates that the integral is performed only when the contact is {\it rolling}; 
i.e., $k_t |\Delta^{t}_{ij}| < \mu_e |{\bf F}^{(n)}_{ij}|$ or $\Delta^{t}_{ij}\cdot v^{t}_{ij} < 0$.
Otherwise, the contact is said to be {\it sliding}.
The expression of the tangential force depends on the state of the contact: 
$\mu_e  |{\bf F}^{(n)}_{ij}| {\bf v}^{(t)}_{ij} / |{\bf v}^{(t)}_{ij}| $ for sliding contact 
and $k_t {\bf \Delta}^{(t)}_{ij}$ for rolling contact.

We consider a bidisperse system in order to avoid crystallization, 
where the diameters of the constituent particles are $0.7d$ and $1.0d$, respectively.
The number of each grain is the same.
For simplicity, we assume that the mass of these grains is the same, which is denoted by $m$.
The dimensions of the system are $L\times L \times H$, 
where we use periodic boundary conditions along the $x$ and the $y$ axes.
In the $z$ direction, there exist two rigid walls that consist of the larger particles.
One of the walls is displaced along the $x$ axis at constant velocity $V$ 
to realize plain shear flow, where the velocity gradient is formed in the $z$ direction.
These grains interact with the bulk grains via the force described above.
This wall is also allowed to move along the $z$ axis 
so that the pressure is kept constant at $P$,  while it is immobile along the $y$ axis.
Namely, the $z$ dimension of the system, denoted by $H$, is a dynamic variable.
The equation of motion of the wall along the $z$ axis is given as 
$M{\ddot H} = F_z - P L^2$, 
where $M$ denotes the mass of the wall and $F_z$ is the repulsive force given by the grains.
We check that the mass of the wall does not influence the result.
Hereafter we set $M=100m$.
We also confirm that the $x$ and $y$ dimensions do not influence the result 
by comparing two systems: $L=10d$ and $20d$.
We thus adopt $L=10d$ to save the computational time.
The traction acting on the moving wall is monitored, which is denoted by $F_x$, 
so that the friction coefficient of the system is defined as $\mu\equiv F_x/L^2/P$.
We do not consider gravity here.

For de-dimensionalization, we set $d=1$, $k=1$, and $m=1$.
The other parameters are chosen as $k_t = k/5$, $\mu_e = 0.6$, and $\zeta = 1$.
The coefficient of restitution vanishes with these parameters, 
which may be justified in modeling rock powder.
However, the general correspondence of these parameter values in DEM 
to the material constants in real life is not rigorous.
Here we estimate the material constants based on the sound velocity and the Young's modulus, 
which are roughly estimated as $d\sqrt{k/m}$ and $k/d$, respectively.
(Note that the numerical factors are neglected.)
Thus, the unit velocity and the unit pressure in DEM are of the order of kilometer per second 
and several tens of Gigapascal, respectively.

\section{Result}
\begin{figure}
\noindent\includegraphics[width=19pc]{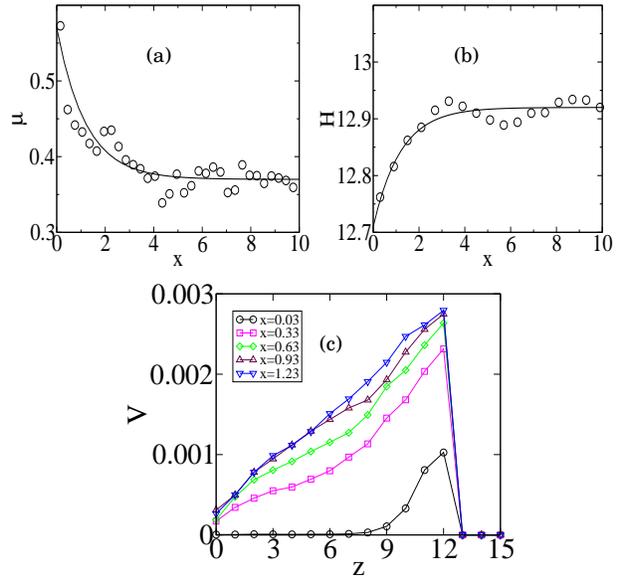}
\caption{\label{relaxation} 
The relaxation behaviors of the system after the velocity change.
Here $V_1=1\times10^{-5}$, $V_2=3\times10^{-3}$, and $P=1\times10^{-3}$.
(a) The friction coefficient $\mu$ and (b) the layer thickness $H$.
The horizontal axes represent the slip distance after the velocity change.
Symbols denote the simulation data, while the solid lines denote the exponential curves, 
Eqs. (\ref{exponentialM}) and (\ref{exponentialH}).
(c) Relaxation of the internal velocity profile, which is defined as 
the instantaneous local mean velocity in the $x$ direction.}
\end{figure}

We first prepare a stationary state with the wall velocity $V_1$.
The stationarity is checked by monitoring the friction coefficient, the volume, 
and the internal velocity gradient.
Then the wall velocity is instantaneously switched from $V_1$ to $V_2$.
The typical response of the system is shown in Figure \ref{relaxation}, where the friction coefficient 
relaxes to a new stationary value corresponding to the new sliding velocity $V_2$.
The sharp increase of the friction coefficient at the instance of the velocity change 
is due to the steeper velocity gradient near the wall.
Then this nonlinear velocity gradient relaxes to the uniform velocity gradient, 
which leads to the relaxation of the friction coefficient.
The relaxation behaviors of the friction coefficient $\mu$ and the layer thickness $H$ 
can be fitted by the exponential curve, as is shown in Figures \ref{relaxation}a and b.
\begin{eqnarray}
\label{exponentialM}
\mu(x) &=& \mu_2 + (\mu_1-\mu_2)\exp(-x/D_c),\\
\label{exponentialH}
H(x) &=& H_2 + (H_1-H_2)\exp(-x/D_c),
\end{eqnarray}
where $x$ denotes the slip distance after the velocity switch and $D_c$ defines the critical slip distance.
We confirm that the critical slip distance is almost the same for the friction coefficient and the layer thickness.
We test several values of $V_1$, $V_2$, $P$, and $H$ to find that 
the choice of $V_1$ does not apparently affect the critical slip distance.
We thus fix $V_1=1\times10^{-5}$ and change $V_2$, $P$, and $H$ in the following.

\begin{figure}
\noindent\includegraphics[width=19pc]{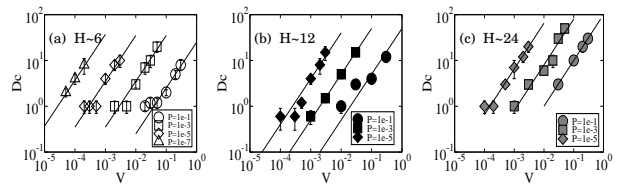}
\caption{\label{V-Dc}
The critical slip distance $D_c$ as a function of the velocity, $V=V_2$.
(a) $H\simeq 6d$, (b) $H\simeq 12d$, and (c) $H\simeq 24d$.}
\end{figure}

As is shown in Figure \ref{V-Dc}, the critical slip distance depends on the slip velocity $V_2$ 
and the normal pressure $P$.
Importantly, the critical slip distance is proportional to the slip velocity so that 
the relaxation time can be defined as the proportional coefficient.
\begin{equation}
\label{tau}
D_c \simeq \tau V_2, 
\end{equation}
where the relaxation time $\tau$ is independent of the velocity (but still depends on the pressure).
In this sense, the relaxation time is more fundamental than the critical slip distance 
in sheared granular layers.
This makes a quite contrast to conventional experiments on friction of two surfaces, 
where the critical slip distance is independent of the sliding velocity (e.g. see \cite{marone2}).
The discrepancy is due to the different physical mechanisms of friction.
In the rubbing of two surfaces, the area of true contact (asperity) determines 
the friction coefficient so that the critical slip distance is of the order of 
the typical dimension of asperities (e.g. tens of micrometers), 
whereas the internal velocity profile determines the friction coefficient in granular layers.

\begin{figure}
\noindent\includegraphics[width=19pc]{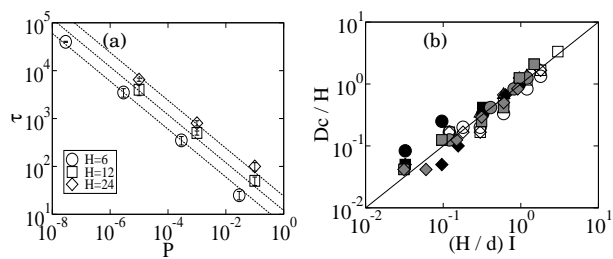}
\caption{\label{fig3}
(a) The pressure dependence of the relaxation time, defined by equation (\ref{tau}).
Each line denotes $\tau/P^{1/2}= 6, 14, 24$.
(b) The critical slip distance $Dc$ divided by $H$ as a function of the inertial number 
$I\equiv V_2/H\sqrt{m/Pd}$ multiplied by $H/d$, i.e., $V_2/\sqrt{m/Pd^3}$.
Thus, this indicates that the critical slip distance is proportional to the layer thickness.
The solid line represents $D_c/H = V_2\sqrt{m/Pd^3}$.
The legends are the same as those in Figure \ref{V-Dc}.}
\end{figure}

Then we discuss the nature of the relaxation time.
In Figure \ref{fig3}a, we find that the relaxation time is inversely proportional to 
the square root of the pressure.
\begin{equation}
\label{sqrtP}
\tau\propto P^{-1/2}.
\end{equation}
Although it has been recognized that the intrinsic time constant in granular matter 
should be scaled as equation (\ref{sqrtP}) from the viewpoint of dimensional analysis, 
this relation has not been confirmed in a dense system.
It is also found that the relaxation time is proportional to the layer thickness, 
as is confirmed in Figure \ref{fig3} a.
\begin{equation}
\label{thickness}
\tau\propto H, 
\end{equation}
which implies that the perturbation propagates into the granular layers at the constant velocity.
This makes a quite contrast to Newtonian fluids, where the velocity field is diffusive 
so that the relaxation time is proportional to the square of the layer thickness.
However, at this point, we cannot derive Eqs. (\ref{sqrtP}) and (\ref{thickness}) 
from the microscopic principle, i.e., the particle dynamics,

From Eqs. (\ref{sqrtP}) and (\ref{thickness}), the relaxation time reads 
\begin{equation}
\label{expression_tau}
\tau = c \frac{H}{d}\sqrt{\frac{m}{Pd}}.
\end{equation}
where $c$ is a numerical factor. We find $c=1.0\pm0.1$ in Figure \ref{fig3}a.
Equivalently, from equation (\ref{tau}), the critical slip distance can be written as 
\begin{equation}
\label{expression_Dc}
D_c = c H V_2 \sqrt{\frac{m}{Pd^3}} = c I\frac{H^2}{d},
\end{equation}
where $I\equiv V_2/H\sqrt{m/Pd}$ is a nondimensional number known as the inertial number 
(see \cite{GDR} for its usefulness in describing granular rheology, 
while it is originally defined by \cite{savage}.)
We can confirm the validity of equation (\ref{expression_Dc}) in Figure \ref{fig3}b.
Note that the critical slip distance is proportional to the layer thickness.
This is consistent with an experiment in which the critical slip distance is scaled with 
the gouge layer thickness \citep{marone1}.

Then we introduce the evolution law for the transient states.
The exponential relaxation of the friction coefficient, equation (\ref{exponentialM}), implies that 
the evolution law is a first order linear differential equation.
Because the critical slip distance depends on the sliding velocity, 
it is more convenient to describe the evolution law with respect to time instead of the slip distance.
The relaxation process is then described by the following evolution law.
\begin{equation}
\label{evolution}
\dot{\mu}(t) = -\tau (\mu (t) - \mu_{\rm ss}),
\end{equation}
where $\mu_{\rm ss}$ is the stationary friction coefficient, which generally depends on 
the inertial number and other nondimensional parameters.
We do not discuss a stationary constitutive law here 
(See, for example, \citep{GDR, dacruz,pouliquen,hatano}).
It is essential to notice that equation (\ref{evolution}) is length-free; 
i.e., the equation does not have any length constants.
In the following section, we discuss some important consequences of this length-free evolution law.

\begin{figure}
\noindent\includegraphics[width=19pc]{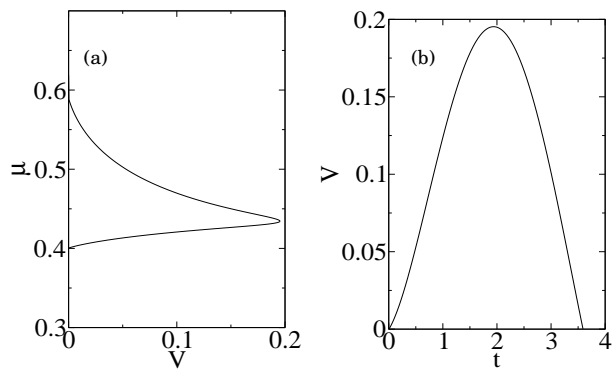}
\caption{\label{sliderdynamics}
Unstable sliding of a spring-block system, described by equations (\ref{slider}) and (\ref{powerlaw}), 
due to the static friction.
(a) Velocity dependence of the friction coefficient during the unstable slip.
(b) Temporal behavior of the block velocity.}
\end{figure}

\section{Discussions}
So far we have discussed the nature of the relaxation time in a transient process of granular layers 
and obtained the evolution law.
In the rest of this paper, we discuss their important consequences in application to a more general situation.
First, using equation (\ref{evolution}), we describe the unstable sliding of a block on granular layers.
The equation of motion is given as
\begin{equation}
\label{slider_raw}
M\ddot{X}=-M\Omega^2 (X-X_0) - \mu N, 
\end{equation}
where $M$ is the mass of a block, $N$ is the normal load, $M\Omega^2$ is the stiffness of the spring, 
and $X_0$ is the equilibrium point of the spring.
As the temporal evolution of the friction coefficient $\mu$ is given by equation (\ref{evolution}), 
the block motion is determined combining these equations.
Before explicitly solving Eqs. (\ref{evolution}) and (\ref{slider_raw}), 
we can see the essential property of the dynamics by dimensional analysis.
Note that equation (\ref{slider_raw}) has the time constant $\Omega^{-1}$ 
and the length constant $N/M\Omega^2$, whereas equation (\ref{evolution}) does not have 
any length constants but the time constant $\tau$.
This means that any lengths defined in the resulting dynamics, such as the critical slip distance, 
is scaled by $N/M\Omega^2$.
Namely, the critical slip distance is not prescribed by the characteristic length of the microscopic geometry 
but more macroscopic parameters: the normal load and the stiffness of the spring.
In order to see this more explicitly, we solve Eqs. (\ref{evolution}) and (\ref{slider_raw}).
To this end, it is convenient to de-dimensionalize equation (\ref{slider_raw}) as 
\begin{equation}
\label{slider}
\Omega^{-2} \ddot{x} = - x - \mu, 
\end{equation}
where $x \equiv (X-X_0) M\Omega^2/N$.
The initial condition of the block motion is given as $x(0)= -\mu(0)$ and $\dot{x} (0)=0$.
We assume that the slip instability is caused by the static friction; 
i.e., $\mu(0) = \mu_s$.
This is easily realizable in granular layers if the initial state is sufficiently consolidated.
Here we consider only $\dot{x}(t)>0$, i.e., we solve the block motion until it stops again.

The dynamics of the block is explicitly solvable if we assume that the dynamic friction coefficient 
is independent of the sliding velocity; i.e., $\mu_{\rm SS} = \mu_d$.
Then the solution is given as 
\begin{eqnarray}
\label{solution1}
x(t) &=& -\frac{\mu_s-\mu_d}{1+C^2} \left[
C\sin\Omega t+\cos\Omega t + C^2 e^{-t/\tau}\right] - \mu_d,\\
\label{solution2}
\mu(t) &=& (\mu_s-\mu_d)e^{-t/\tau} + \mu_d,
\end{eqnarray}
where $C\equiv\Omega\tau$ is the ratio of the two time constants in Eqs. (\ref{evolution}) and (\ref{slider}).
Then the the critical slip distance is apparent from Eqs. (\ref{solution1}) and (\ref{solution2}).
\begin{equation}
D_c\simeq[x(\tau)+\mu_s]\frac{N}{M\Omega^2}.
\end{equation}
The numerical factor $[x(\tau)+\mu_s]$ is of the order of $1$ unless $C\ll 1$.
Note that $D_c$ is scaled with $N/M\Omega^2$, which is approximately equal to the slip distance.
We wish to stress that, in granular layers, the critical slip distance is not scaled with 
the characteristic length of the microscopic geometry such as the surface roughness.
This is the natural consequence of the legth-free nature of the relaxation, 
represented in the form of equation (\ref{tau}).

We also test a more plausible law for the dynamic friction, which is recently found in the DEM simulation.
\begin{equation}
\label{powerlaw}
\mu_{\rm SS}=\mu_0+sI^{\phi}, 
\end{equation}
where $\phi\simeq0.3$ and $s$ is a numerical factor of the order of $0.1$ \citep{hatano}.
Here we interpret $I$ as $V\sqrt{m/Nd}$.
The resulting dynamics is shown in Figure \ref{sliderdynamics}, 
where we adopt $\mu_0=0.3$, $s=0.2$, and $C=1$.
The dynamics is very similar to that obtained in an experiment \citep{nasuno},
in which the nondimensional number $C$ is estimated to be of the order of $1$.

Despite the feasibility in reproducing an experimental result on granular matter, 
we have to remark that the nondimensional parameter $C=\Omega\tau$ 
may be very small under a seismogenic condition.
For example, if we assume that $P=100$ MPa, $d=10$ $\mu$m, and $H=1$ cm, 
using equation (\ref{expression_tau}), the relaxation time $\tau$ is of the order of $10^{-4}$ s.
As the seismic slip takes place in seconds, $C$ is of the order of $10^{-4}$ 
so that the critical slip distance is negligible compared with the total slip distance.
However, note that the framework of the length-free evolution law is not limited to 
the relaxation process of the velocity profile.
It is straightforward to extend the present evolution law to incorporate any processes 
that affect the frictional strength. 
One of the most illustrating examples is a mechanochemical effect such as 
thermal decomposition of calcite, the rate constant of which is on the order of 
$1$ sec or even much larger depending on the temperature \citep{hirono}.
A plausible modeling is in progress and the result will be presented elsewhere.


\begin{acknowledgments}
The author gratefully acknowledges the discussion with Osamu Kuwano, 
Ryosuke Ando, Jean-Pierre Villote, and Pascal Bernard.
\end{acknowledgments}

%




%
%

\end{article}




\begin{thebibliography}{99}
\bibitem[{\textit{Aharonov and Sparks}(2002)}]{aharonov}
Aharonov, E. and D. Sparks (2002), 
Shear profiles and localization in simulations of granular materials,
{\it Phys. Rev. E} {\textit 65}, doi:051302, 10.1103/PhysRevE.65.051302.


\bibitem[{\textit{Aki}(1987)}]{aki}
Aki, K. (1987), 
Magnitude-frequency relation for small earthquakes: A clue to the origin of fmax of large earthquakes,
{\it J. Geophys. Res.} {\textit 92}, 1349-1355, doi:10.1029/JB092iB02p01349.


\bibitem[{\textit{Cundall and Strack} (1979)}]{cundall}
Cundall, P. A. and O. D. L. Strack (1979), 
A distinct element model for granular assemblies. Geotechnique,
{\it Geotechnique} {\textit 29}, 47--65.

\bibitem[{\textit{da Cruz et al.} (2005)}]{dacruz}
da Cruz, F., S. Emam, M. Prochnow, J. N. Roux, and F. Chevoir (2005), 
Rheophysics of dense granular materials: Discrete simulation of plane shear flows,
{\it Phys. Rev. E} {\textit 72}, 021309, doi:10.1103/PhysRevE.72.021309.

\bibitem[{\textit{Dieterich}(1979)}]{dieterich}
Dieterich, J. H. (1979), 
Modeling of rock friction 1. Experimental results and constitutive equations,
{\it J. Geophys. Res.} {\textit 84(B5)}, 2161--2168, doi:10.1029/JB084iB05p02161.

\bibitem[{\textit{Engelder}(1974)}]{engelder}
Engelder, J. T. (1974), Cataclasis and the Generation of Fault Gouge, 
{\it Bull. Geological Soc. Am.} {\textit 85}, 1515--1522, 
doi:10.1130/0016-7606(1974)85.

\bibitem[{\textit{Garz{\'o} and Dufty}(1999)}]{garzo}
Garz{\' o}, V. and J. W. Dufty (1999), 
Dense fluid transport for inelastic hard spheres,
{\it Phys. Rev. E} {\textit 59}, 5895--5911, doi:10.1103/PhysRevE.59.5895.

\bibitem[{\textit{GDR MiDi}(2004)}]{GDR}
GDR MiDi (2004), On dense granular flows,
{\it Euro. Phys. J. E} {\textit 14}, 367--371, doi:10.1140/epje/i2003-10153-0.

\bibitem[{\textit{Hatano}(2007)}]{hatano}
Hatano, T. (2007), 
Power-law friction in closely packed granular materials,
{\it Phys. Rev. E.} {\textit 75}, 060301(R), doi:10.1103/PhysRevE.75.060301.

\bibitem[{\textit{Hazzard and Mair}(2003)}]{mair}
Hazzard, J.F. and K. Mair (2003), 
The importance of the third dimension in granular shear,
Geophys. Res. Lett. {\textit 30}, 1708, doi:10.1029/2003GL017534.

\bibitem[{\textit Hirono et al. (2007)}]{hirono}
Hirono, T., T. Yokoyama, Y. Hamada, W. Tanikawa, T. Mishima, M. Ikehara, V. Famin, 
M. Tanimizu, W. Lin, W. Soh, and S. Song (2007), 
A chemical kinetic approach to estimate dynamic shear stress during 
the 1999 Taiwan Chi-Chi earthquake, 
{\it Geophys. Res. Lett.} {\textit 34}, L19308, doi: 10.1029/2007GL030743.


\bibitem[{\textit{Ida}(1973)}]{ida}
Ida, Y. (1973), The maximum acceleration of strong ground motion,
{\it Bull. Seism. Soc. Am.} {\textit 63}, 959-968.

\bibitem[{\textit{Ide and Takeo}(1997)}]{ide}
Ide, S., and M. Takeo (1997), 
Determination of constitutive relations of fault slip based on seismic wave analysis, 
{\it J. Geophys. Res.} {\textit 102(B12)}, 27379--27391, 10.1029/97JB02675.


\bibitem[{\textit{Jop et al.}(2004)}]{pouliquen}
Jop, P., Y. Forterre, and O. Pouliquen (2006),  
A constitutive law for dense granular flows,
{\it Nature}, {\textit 441}, 727--730, doi:10.1038/nature04801.

\bibitem[{\textit{Marone and Kilgore}(1998)}]{marone1}
Marone, C. and B. Kilgore (1993), 
Scaling of the critical slip distance for seismic faulting with shear strain in fault zones, 
{\it Nature} {\textit 362}, 618--621, doi:10.1038/362618a0.

\bibitem[{\textit{Marone}(1998)}]{marone2}
Marone, C. (1998), 
Laboratory-derived friction laws and their application to seismic faulting,
{\it Ann. Rev. Earth Planet. Sci.} {\textit 26}, 643--696, doi: 10.1146/annurev.earth.26.1.643.

\bibitem[{\textit{Nasuno et al.}(1998)}]{nasuno}
Nasuno, S., A. Kudrolli, A. Bak, and J. P. Gollub (1998), 
Time-resolved studies of stick-slip friction in sheared granular layers,
{\it Phys. Rev. E.} {\textit 58},  2161--2171, doi:10.1103/PhysRevE.58.2161.

\bibitem[{\textit{Ohnaka}(2000)}]{ohnaka1}
Ohnaka, M., 2000, A physical scaling relation between the size of an earthquake and its nucleation zone size, 
{\it Pure and Applied Geophysics} {\textit 157}, 2259--2282, doi: 10.1007/PL00001084.

\bibitem[{\textit{Ohnaka and Shen}(1999)}]{ohnaka2}
Ohnaka, M., and L.-f. Shen (1999), Scaling of the shear rupture process from nucleation to dynamic propagation: Implications of geometric irregularity of the rupturing surfaces, 
{\it J. Geophys. Res.} {\textit 104}, 817--844, doi:10.1029/1998JB900007.

\bibitem[{\textit{Ohnaka}(2003)}]{ohnaka3}
Ohnaka, M. (2003), A constitutive scaling law and a unified comprehension for frictional slip failure, 
shear fracture of intact rock, and earthquake rupture, 
{\it J. Geophys. Res.} {\bf 108(B2)}, 2080, doi:10.1029/2000JB000123.

\bibitem[{\textit{Savage and Hutter}(1989)}]{savage}
Savage, S. B. and K. Hutter (1989), 
The motion of a finite mass of granular material down a rough incline, 
{\it J. Fluid. Mech.} {\textit 199}, 177--215, doi:10.1017/S0022112089000340.

\bibitem[{\textit{Scholz}(1987)}]{scholz}
Scholz, C. H. (1987), Wear and gouge formation in brittle faulting, 
{\it Geology} {\textit 15}, 493--495, doi:10.1130/0091-7613(1987)15.

\bibitem[{\textit{Scholz}(2002)}]{scholtz}
Scholz, C. H. (2002),
{\it The mechanics of earthquakes and faulting}, (Cambridge University Press, Cambridge).
\end{thebibliography}
\end{document}